\documentclass{emulateapj}
\usepackage{apjfonts}
\usepackage{latexsym}
\usepackage{color}

\def\ltsim{ \,{}^<_\sim\, }

\shorttitle{RR Lyrae and the dSph--Halo connection}
\shortauthors{Fiorentino et al.}
\begin{document}
\def\gsim{\;\lower.6ex\hbox{$\sim$}\kern-6.7pt\raise.4ex\hbox{$>$}\;}
\def\lsim{\;\lower.6ex\hbox{$\sim$}\kern-6.7pt\raise.4ex\hbox{$<$}\;}
\title{Weak Galactic halo--dwarf spheroidal connection from RR Lyrae stars}

\author{Giuliana Fiorentino$^{1}$, Giuseppe Bono$^{2,3}$, Matteo Monelli$^{4,5}$, Peter B. Stetson$^{6}$, Eline Tolstoy$^{7}$, Carme Gallart$^{4,5}$, Maurizio Salaris$^{8}$, Clara Mart{\'i}nez-V{\'a}zquez$^{4,5}$ \& Edouard~J.~Bernard$^{9}$}
\email{giuliana.fiorentino@oabo.inaf.it}
\affil{$^{1}$INAF-Osservatorio Astronomico di Bologna, via Ranzani 1, 40127, Bologna.}
\affil{$^{2}$Dipartimento di Fisica, Universit\'{a} di Roma Tor Vergata, Via della Ricerca Scientifica 1, 00133 Roma, Italy.}
\affil{$^{3}$INAF-Osservatorio Astronomico di Roma, Via Frascati 33, 00040 Monte Porzio Catone, Italy}
\affil{$^{4}$Instituto de Astrof\'{i}sica de Canarias, Calle Via Lactea s/n, E38205 La Laguna, Tenerife, Spain.}
\affil{$^{5}$Departmento de Astrof\'{i}sica, Universidad de La Laguna, E38200 La Laguna, Tenerife, Spain.}
\affil{$^{6}$National Research Council, 5071 West Saanich Road, Victoria, BC V9E 2E7, Canada.}
\affil{$^{7}$Kapteyn Astronomical Institute, University of Groningen, Postbus 800, 9700 AV Groningen, The Netherlands}
\affil{$^{8}$Astrophysics Research Institute, Liverpool John Moores University IC2, Liverpool Science Park 146 Brownlow Hill Liverpool L35RF, UK.}
\affil{$^{9}$SUPA, Institute for Astronomy, University of Edinburgh, Royal Observatory, Blackford Hill, Edinburgh EH9 3HJ, UK}
\begin{abstract}

We discuss the role that dwarf galaxies may have played in the
formation of the Galactic halo (Halo) using RR Lyrae stars (RRL)
as tracers of their ancient stellar component. The comparison is
performed using two observables (periods, luminosity amplitudes)
that are reddening and distance independent.  
Fundamental mode RRL in six dwarf spheroidals and eleven ultra
faint dwarf galaxies ($\sim$1,300) show a Gaussian period
distribution well peaked around a mean period of
$<$Pab$>=$0.610$\pm$0.001 days ($\sigma$=0.03). The Halo RRL
($\sim$15,000) are characterized by a broader period
distribution.\par
The fundamental mode RRL in all the dwarf spheroidals apart from
Sagittarius are completely lacking in High Amplitude Short Period
(HASP) variables, defined as those having P$\ltsim$0.48
days and A$_V\ge$0.75mag. Such variables are not uncommon in
the Halo and among the globular clusters and massive dwarf 
irregulars. To further interpret this evidence, we considered eighteen
globulars covering a broad range in metallicity
(-2.3$\lsim$[Fe/H]$\lsim$-1.1) and hosting more than 35 RRL
each.\par  
The metallicity turns out to be the main parameter, since only
globulars more metal--rich than [Fe/H]$\sim$-1.5 host RRL in the
HASP region. This finding suggests that dSphs similar to
the surviving ones do not appear to be the major building--blocks
of the Halo. Leading physical arguments suggest an {\it extreme}
upper limit of $\sim50\%$ to their contribution. On the other
hand, massive dwarfs hosting an old population with a broad
metallicity distribution (Large Magellanic Cloud, Sagittarius) may
have played a primary role in the formation of the Halo.  
\end{abstract}

\section{Introduction}\label{intro} 
\begin{figure*}
\centering
\includegraphics[width=15.5cm]{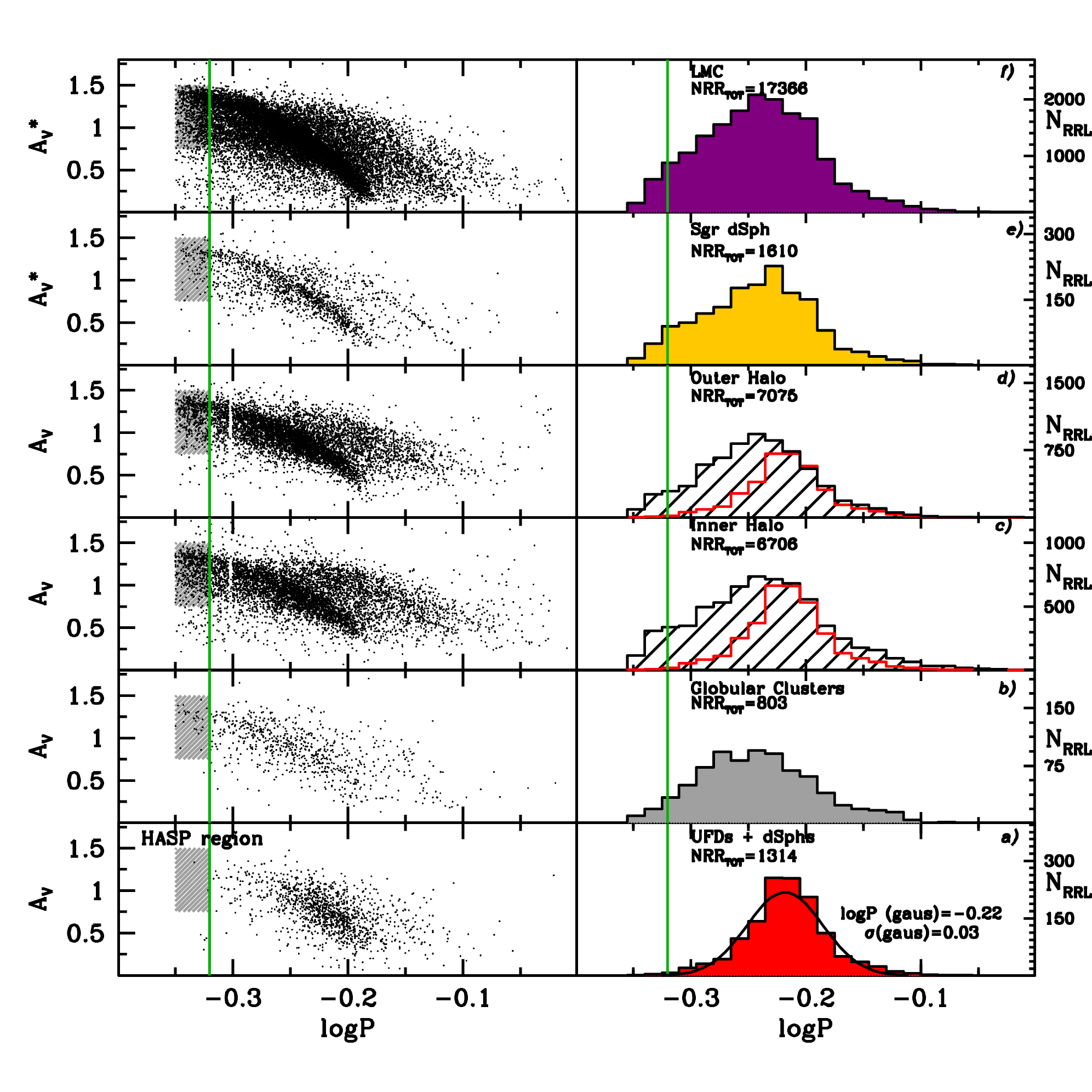}
\caption{{\it Left --}  Bailey diagram, Period vs V--Amplitude
distributions for RRab observed in the LMC ({\it f} panel), in Sgr
dSph ({\it e} panel), the Halo with log P $\ge$-0.35 (top panels
{\it c} and {\it d}), in GGCs ({\it b} panel), in six dSphs plus
eleven UFDs ({\it a} panel). A$_V^*$ is derived from A$_I$ using
the A$_I$/A$_V$=0.63 \citep{dicriscienzo11}. The green solid
vertical line represents the minimum period observed for RRab in
spheroidals, i.e. P $\sim$0.48 d. We have highlighted the region
where HASP RRab are missing (grey); {\it Right --} Period
histograms for the same sample of RRab. Red histograms in panels
{\it c} and {\it d} represent the UFD$+$dSphs distribution
rescaled to the Halo ones as explained in the text.  
\label{fig1}}
\end{figure*}

The early suggestion by \cite{searle78} that the Milky Way (MW) outer 
Halo formed from the aggregation of protogalactic fragments  was 
supported i) theoretically, by $\Lambda$CDM simulations of galaxy 
formation in which small galaxies form first and then cluster to 
form larger galaxies, and ii) observationally,  
by the discovery of stellar streams and merging satellites in the 
MW \citep[][]{ibata94} and in other galaxies. However, the 
characteristics of the halo building--blocks is still a matter 
of debate. In particular, the
question of whether the current dwarf spheroidal (dSph) satellites
of the MW are surviving representatives of the Halo's
building--blocks has been explored in several works 
\citep[][]{tolstoy09}. The conclusions of these works differ in
some details, but they suggest that there are difficulties in 
forming the Halo exclusively with dwarfs similar to
current MW satellites. Among these, the studies using element 
ratios \citep{venn04}, of stellar populations in dSphs and in 
the Halo are based on tracers (RGs) covering a wide range in age. 
Thus, they suffer from the drawback that some of the dSph present 
a complex evolution spanning several Gyr,
while the evolution of the halo building--blocks was likely
interrupted at early times, when they were accreted into the halo.
This is the reason why old stellar tracers are crucial in the 
comparison between the MW halo and dSphs.  

A real possibility to isolate the ancient (age $\ge$10Gyr)
populations in these different stellar systems is offered by a
special class of low mass, radial variables: RR Lyrae stars (RRLs). 
They pulsate in the fundamental (RRab) and in the first 
overtone mode (RRc). Due to their variability and relatively distinct light 
curves RRLs can be easily distinguished from other stars.
Extensive variability surveys of our
Galaxy have been performed and are nowadays releasing their final
catalogues. We have compiled a huge catalogue ($\sim$14,700 stars)
from QUEST \citep{vivas04,zinn14}, ASAS \citep{szczygiel09} and
CATALINA \citep[][]{drake13} surveys that have classified RRLs and provided Johnson $V$--band 
magnitudes and amplitudes.
The final catalog is mainly based on CATALINA RRLs (85\%) and 
covers a large range in galactocentric distances (5Kpc$\lsim$d$_G\lsim$60Kpc). 
Moreover, the sample radial distribution does not show evidence of gaps.
This makes possible a direct comparison with dSphs where RRLs 
are always observed. We have gathered the results from accurate and quite
complete photometric studies of classical dSphs (Draco,
Carina, Tucana, Sculptor, Cetus and Leo~I and some ultra faint
dwarfs, hereinafter UFDs) that different research groups have
carried out during the last ten years \citep[see][]{stetson14b}.  

In \citet{stetson14b} we performed a first detailed analysis of
the RRLs properties using these sizable samples.  Very
interestingly, comparing their period--amplitude (or Bailey)
diagrams, we highlighted that, in the sample of six dSphs plus
eleven UFDs that we considered, there are {\it no\/} RRab stars
with A$_V\ge$0.75mag and P $\ltsim$0.48d. This was first found by
\citet{bersier02} in Fornax and not explained by the temporal
sampling of the observations, since their probability to detect a
period in such a range was always higher than 66\%. The authors
attributed it instead to the transition period between RRab and
RRc--type variables. The same applies to the Draco dSph as
discussed by \citet{catelan09}. We have observed this evidence in
another five dSphs and eleven UFDs and named it the missing High
Amplitude Short Period (HASP) RRab problem in dSphs. This
evidence can not be related to photometric incompleteness of a
single photometric dataset, particularly since high amplitude RRLs
are the easiest to recognize among the variable candidates.  

In this letter, using the properties of RRab in GGCs and taking
advantage of predictions from theoretical models, we propose an
explanation for the missing HASP problem in dwarfs. We
also give a rough estimate of the upper limit to the contribution
of dSph-like galaxies to the Halo stellar population. We close the
letter extending the discussion to the possible contribution to
the Halo of systems similar to the Large Magellanic Cloud (LMC)
and the Sagittarius (Sgr) dwarfs.  

\begin{deluxetable*}{ccccccc}
\centering
\tabletypesize{\tiny}
\tablewidth{0pt}
\tablenum{1}
\tablecaption{Mean RRab properties.}
\label{tabper}
\tablehead{
 \colhead{}   &
 \colhead{Gal.In}   &
 \colhead{Gal.Out}  &
 \colhead{dSphs}      &
 \colhead{LMC}       &
 \colhead{Sgr}       & 
 \colhead{GGCs}       
\\
 \colhead{}       &
 \colhead{halo}   &
 \colhead{halo}   &
 \colhead{+UFDs}       &
 \colhead{}       &
 \colhead{}       &
 \colhead{}   
}
\startdata
$<$Pab$>$  & 0.584$\pm$0.001 & 0.576$\pm$0.001 & 0.610$\pm$0.001 & 0.575$\pm$0.002 & 0.576$\pm$0.001 & 0.580$\pm$0.002\\
$\sigma_{Pab}$   & 0.08  & 0.07 & {\bf 0.05} & 0.07 & 0.07 & 0.07 
\enddata 
\end{deluxetable*}

\section{The missing HASP RRab in dSphs}\label{hasp} 

Fig.~\ref{fig1} shows the Bailey diagrams (left) and  the period
distributions (right) for the RRab observed in dSphs ({\it a}
panel), in GGCs ({\it b} panel), in the inner (d$_G
\lsim$16\footnote{ 
This distance is slightly smaller than the distance adopted
by \citealt{stetson14b} (14Kpc). The difference does not affect the 
conclusions of the paper.} Kpc,
{\it c} panel) and in the outer (d$_G \gsim$16 Kpc, {\it d} panel)
halo. The boundary between inner and outer halo has been
arbitrarily chosen to have comparable numbers of RRLs in the two
samples and taking into consideration the value found in
\citet[][d$_G \sim$15--20 Kpc]{carollo07}. We take advantage also of
the exceptionally complete LMC sample from OGLE III \citep[{\it f}
panel,][]{soszynski10b} and of the recent release of OGLE IV that
includes Sgr \citep[{\it e} panel,][]{soszynski14}. The shaded
grey area shows the location in Bailey diagram of the HASP region. 
Our analysis focuses on fundamental RRLs ($\log P\ge$-0.35).
The shorter period first overtones will not be included, because
they have smaller luminosity amplitudes and are, at fixed limiting 
magnitude, more affected than RRab by completeness problems.  

The period distributions of RRab stars plotted in the right panels
of Fig.~\ref{fig1} are quite different even if the mean periods
achieve similar values to within 1$\sigma$ (Table~\ref{tabper}). 
In particular, the shape of the histogram corresponding to dSph
and UFD is strikingly different from the rest. Not only the HASP 
are absent (except for two in Cetus, that appear to be
peculiar for other reasons), but there is also a dearth (or a
lower fraction) of short period variables (P$\lesssim$-0.25),
compared to the other five samples. The symmetry of the dSphs 
period distribution can be fitted with a Gaussian function ($\sigma=$0.03, see Fig.~\ref{fig1}) and suggests also that metal--poor RRab in UFDs
are still a minor fraction of the entire sample
(NRR$_{UFD}$/NRR$_{UFD+dSphs}=$3\%). Indeed, the RRab in UFDs
tend to contribute significantly to the long period tail of dSphs
\citep[see Fig.~9 in][]{stetson14b}. The mean and the $\sigma$ of the period distribution
observed in dSphs (in bold in Table~\ref{tabper}) does not
increase when the RRab sample is almost doubled thanks to the
inclusion of newly detected RRab stars (PBS, priv. comm.) in
Fornax ($\gsim$1000 RRab) and Sculptor ($\sim$300). 

\begin{figure*}
\includegraphics[width=15.5cm]{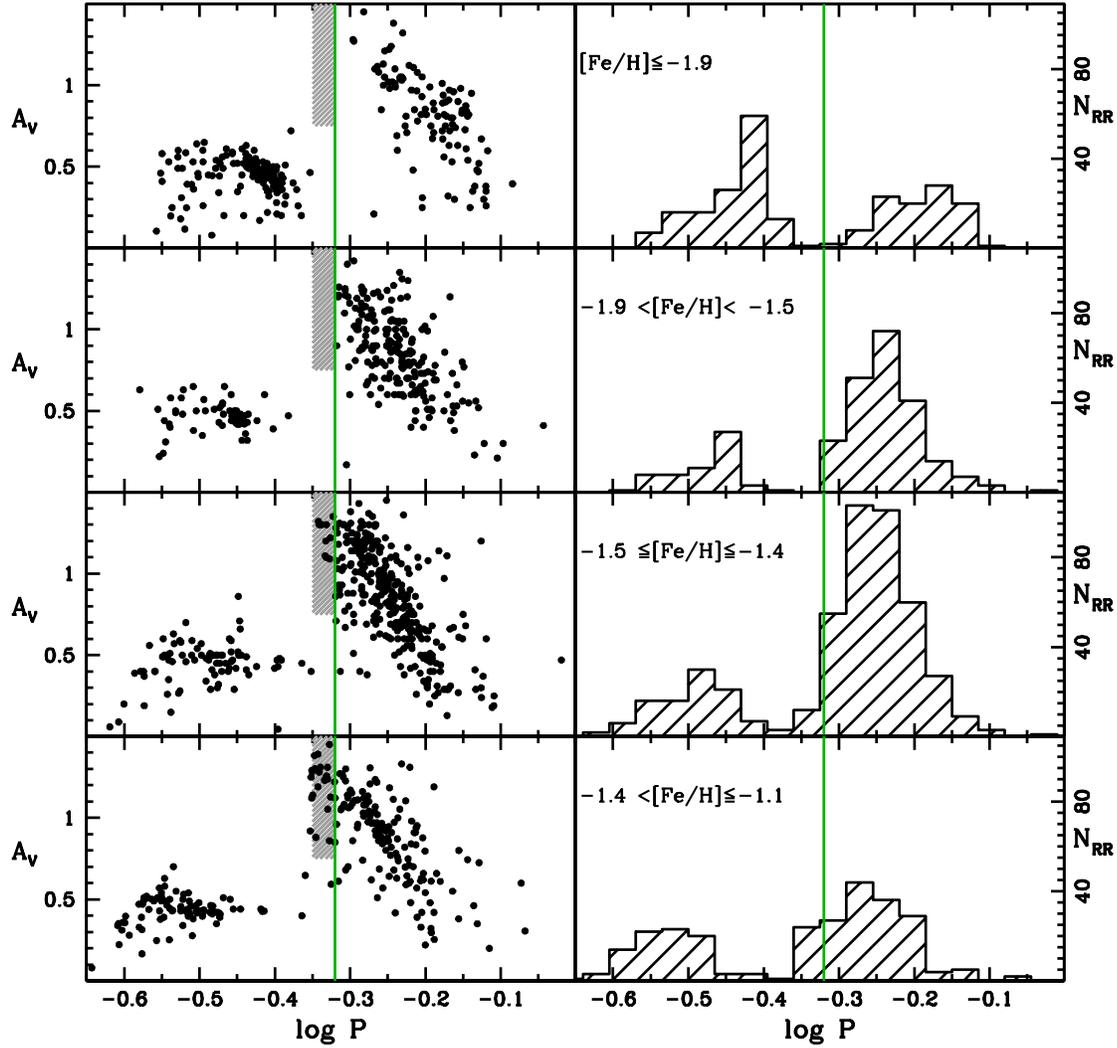}
\caption{Same than in Fig.~\ref{fig1} but for two LMC and sixteen
Galactic GCs. They are grouped accordingly to \citet{harris96} in metallicity bins increasing from top to bottom
panels. {\it First bin)} NGC~7078, NGC~4590, NGC~5024, NGC~2257;
{\it second bin)} NGC~1466, NGC~5286, NGC~7006, NGC~3201, IC4499;
{\it third bin)} NGC~6715, NGC~6934, NGC~6229, NGC~6981, NGC~5272;
{\it fourth bin)}  NGC~5904, NGC~6266, NGC~6121, NGC~6362.  Most
of the RRLs used in GCs belong to the catalog of \citet{clement01} with few exceptions, namely NGC~6121 \citep[or
M4,][]{stetson14a}, NGC~6934\citep{kaluzny01}, NGC~7006\citep{wehlau99}, NGC~2257\citep{walker93} and NGC~1466\citep{walker92}.  
\label{fig2}}
\end{figure*}

\subsection{Why HASP RRLs are missing in dSph?} 

Globular Clusters (GCs) are fundamental laboratories to constrain
old stellar populations, 
since individually they host stars with similar age and 
chemical composition.
To investigate the fine structure of the Instability Strips (IS) we
selected 16 Galactic GCs hosting at least 35 RRLs according to the \citet{clement01} catalog (2013 edition, see
Fig.~\ref{fig2}). To extend the metallicity range
covered by the selected GGCs we included two LMC globulars, namely
NGC~1466 and NGC~2257 \citep{walker92,walker93}. The entire sample
covers a range in metallicity of [Fe/H] from $\sim$--2.3 to --1.1
\citep[][2010 edition]{harris96}. To constrain the metallicity
dependence the entire sample of GCs was split into four arbitrary
metallicity bins. Every bin in metallicity includes at least four
GCs. 

A glance at the data plotted in Fig~\ref{fig2} clearly shows that the
HASP region starts to be filled only when RRLs have a
metallicity $\gtrsim$--1.5dex. It becomes more populated when the
metal content increases to --1dex. The above evidence suggests
the hypothesis that metallicity is the key parameter causing the
lack of HASPs in dSphs. It would imply that the maximum
metallicity reached by the stellar population to which the RRL
belong in dSphs is [Fe/H]$\lesssim$--1.5dex, and that the other
stellar systems have reached a higher metallicity at the early
time when they were still able to produce stars of a mass suitable
for becoming today's RRL. Additionally, a firm dependence of the
{\it mean\/} periods on the metallicity can be observed in
Figure~\ref{fig2}. The $\langle$Pab$\rangle$, when moving from the
metal--poor to the metal--rich regime, decreases from
0.644$\pm$0.007 to 0.599$\pm$0.006d, while the
$\langle$Pc$\rangle$ decreases from 0.364$\pm$0.003 to
0.300$\pm$0.004d. Moreover and even more
importantly, the fraction of HASPs over the total number
of RRab stars is vanishing in the two most metal--poor bins and
becomes of the order of 3\% and 14\% in the two most metal--rich
bins, in order of increasing metallicity. 

\begin{deluxetable*}{cccccccccccc}
\centering
\tabletypesize{\tiny}
\tablewidth{0pt}
\tablenum{2}
\tablecaption{HASP fractions.}
\label{tabHASP}
\tablehead{
 \colhead{}   &
 \colhead{Gal. In$^{1}$}   &
 \colhead{Gal. Out$^{1}$}   &
 \colhead{dSphs$^{2}$} &
 \colhead{LMC$^{3}$}  &
 \colhead{SMC$^{4}$}  &
 \colhead{M31$^{5}$} &
 \colhead{M31$^{6}$} & 
 \colhead{M32$^{7}$}  & 
 \colhead{M33$^{8}$}  & 
 \colhead{Sgr$^{9}$}  & 
 \colhead{Gal.$^{9}$}   
\\
 \colhead{}   &
 \colhead{Halo}   &
 \colhead{Halo}   &
 \colhead{+UFDs} &
 \colhead{}  &
 \colhead{}  &
 \colhead{halo} &
 \colhead{field} & 
 \colhead{}  & 
 \colhead{}  & 
 \colhead{}  & 
 \colhead{Bulge}   
}
\startdata
N$_{HASP}$/NRRL&8\% & 6\% & 0\% & 6\% & 1\% & 3\% & 9--12\%$^a$ & 7\%$^a$ & 2\%$^a$ & 6\% & 17\% \\ 
<[Fe/H]>$\pm$$\sigma$ &n.c.& n.c.& $\lsim$-1.4$\pm$0.1$^{11}$ & -0.5$\pm$0.4$^{11}$ & -1.0$\pm$0.2$^{11}$ & n.c. & n.c.& -0.25$\pm$0.1$^{11}$ &-0.5$\pm$0.1$^{12}$ &-0.4$\pm$0.2$^{11}$ &-0.3$\pm$0.5$^{13}$\\
\enddata 
\tablecomments{$^a$ These numbers should be cautiously treated due
to the limited temporal sampling of the data used to identify and
characterize RRLs.\\ {\it References}$^{1}$ Compilation in this
paper;$^{2}$ Compilation made in \citet{stetson14b};$^{3}$\citet{soszynski10b};$^{4}$\citet{soszynski09a};$^{5}$
\citet{brown04,jeffery11,bernard12};$^{6}$\citet{jeffery11};$^{7}$\citet{fiorentino12a};$^{8}$\citet{yang10};$^{9-10}$\citep{soszynski14}; $^{11}$\citep{mcconnachie12};
$^{12}$\citep{bresolin10}; $^{13}$\citep{uttenthaler12}.} 
\end{deluxetable*}

To further validate the above trend we investigated the occurrence
of HASPs in other nearby stellar systems (see Table~\ref{tabHASP}). 
For some of these systems,
space observations (in F606W filter) are available. In order to
select the HASP RRab we converted F606W amplitudes into
the Johnson--Cousin photometric system, assuming
A$_{F606W}/$A$_V$=0.92 \citep{brown04}.  We selected RRab with
periods shorter than 0.48d and luminosity amplitudes larger 
than A$_{F606W}=$0.69. We found that the
ratio of HASP RRL to total number of RRab follows a trend
similar to GCs, and indeed, they range from a few percent in
systems where the mean metallicity is poor (SMC) to more
than $\sim$10\% in more metal--rich systems (Bulge).  

In this context, the two peculiar RRab in Cetus \citep{bernard09},
located in the HASPs region, might trace the tail of a
metal--rich stellar component. Their luminosity is $\sim$0.1mag
fainter than the remaining $\sim$500 RRLs, thus suggesting an
important metallicity increase in the early star formation  event
experienced by this quite massive galaxy.  

\subsection{Insights from pulsation and evolutionary theory}\label{theory}

Non-linear, convective hydrodynamical models of radial variables 
indicate that RRab have their largest amplitudes close to the 
fundamental blue edge \citep[FBE,][]{bono94}.

The FBE boundary is almost constant over a broad range of metal
abundances (--2.3$\lsim$[Fe/H]$\lsim$--1.3, \citet{bono95a}). This
means that the pulsation properties of an RRL across the IS are
dictated mostly by its evolution. A change in chemical composition
causes a change in stellar mass and in luminosity, and in turn a
change in the morphology of the evolutionary paths crossing the
IS.  Pulsation and evolutionary prescriptions indicate that the
minimum period reached by RRab, i.e., the period at the FBE,
decreases as the metal content increases. In particular,
\citet{bono97d} showed that logP$_{ab}^{min}$ decreases from --0.26
to --0.37 when Z increases from 0.0001([Fe/H]$\sim$--2.3, using
$\alpha$--enahnced values) to 0.001([Fe/H]$\sim$--1.3). These
predictions agree quite well with the minimum period observed in
M3 and in M15---globulars characterized by sizable samples of RRLs.  

This scenario suggests that old stars in dSphs, in spite of their
complex star formation and chemical enrichment history, are
characterized by a narrow metallicity distribution when compared
with relatively ``simple'' stellar systems in the MW such as
GCs. A complementary conclusion was reached by \citet{salaris13}
studing in detail the Horizonthal Branch morphology of the
Sculptor dSph. They found that this can be explained, at odds with
GCs, without invoking He--enhanced models \citep[][]{rood73}.
This evidence further supports the above findings, since an
increase in the helium content would imply, at fixed intrinsic
parameters, a steady increase in the pulsation period
\citep{marconi11}, thus further exacerbating the HASPs
problem.    

\begin{figure}
\includegraphics[width=8.5cm]{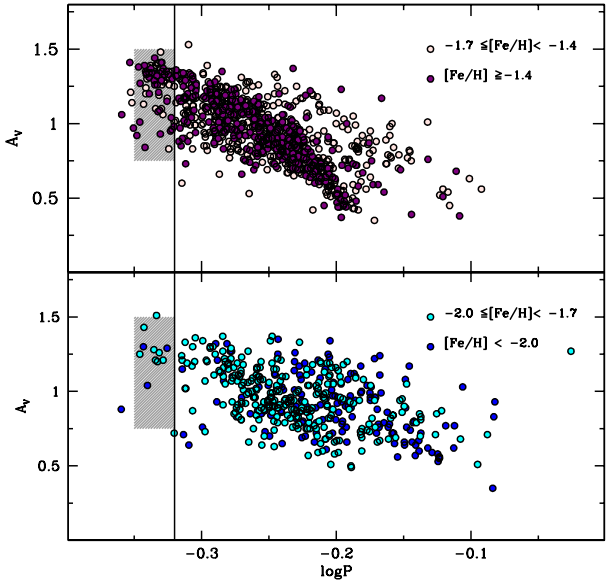}
\caption{Same as in Fig.~\ref{fig1}, but for the RRLs for 
which an estimate of metallicity exists from SDSS \citep{drake13}.  
\label{fig3}}
\end{figure}

\subsection{Evidence from SDSS optical spectra}\label{spectra}

The recent evidence of multiple stellar
populations in GCs \citep[][and reference therein]{monelli13}
and the fact that the horizontal branch in
dSphs can be reproduced without assuming any helium enrichment or
complex mass--loss law \citep{salaris13}, induce us to question
whether we are using too complex stellar systems to understand the
HASP dearth in dSph. In order to corroborate the
hypothesis that low metallicity is the cause of the missing HASP 
in dSphs, we take advantage of the medium resolution SDSS
spectra available for a sample of $\sim$1,400 fundamental mode
RRLs \citep{drake13}. Given that the RRLs span a magnitude range
from 14 and 20mag, the errorbar of each individual metallicity
estimate has been estimated of $\lsim$0.2dex as discussed in
\citep{yanny09}. In Fig.~\ref{fig3} we show the Bailey
distribution of these stars grouped in four metallicity bins.
Note that RRLs in the above metallicity bins cover similar 
ranges in Galactocentric distances (5$\lsim$d$_G\lsim60$ Kpc).
We can clearly see that metal--rich groups of RRLs ([Fe/H]>--1.7, top
panel) tend to populate the HASP region whereas the
metal--poor ones ([Fe/H]$\lsim$--1.7) leave this region almost
empty, starting from period lpgP$\lsim$--0.3. Although the
absolute value of the metallicity at which the transition occurs
would need a careful calibration of these data, we find a trend in
agreement with that observed in GCs.

\section{Implications for the early formation of the Halo}

We have shown that the period distribution of RRL in nearby dSphs
follows a Gaussian distribution with a smaller dispersion
($\sigma=$0.03) than the Halo one, which is more skewed to short
periods. This peculiarity of the dSphs is not observed in GGCs, the LMC or the
Sgr dwarf. Furthermore, we found evidence that, in order to
populate the HASP region, an old component more metal--rich
than [Fe/H]$\sim$--1.5 is required (see Fig.~\ref{fig2}). In the
following, we will analyze the evidence provided by the RRL
populations on the building of the Halo from the combination of
different types of progenitors. In this exercise, we will assume
that the Halo sample is statistically significant, i.e., that an
increase in its size would not affect the shape of the period
distribution\footnote{The validity of this assumption depends on
the Galactocentric distance. The new variability surveys
\citep[CATALINA,][their Fig.~13]{drake13} appear to be quite
complete ($\sim$50\%) out to 40 kpc (V$\sim$18mag). If this
assumption is wrong---that is, if the Halo is affected by a
significant amount of uncompleteness---this could affect the
results of the Kolmogorov--Smirnov test presented in this
section. However, we can be confident that it would not affect
the HASP fractions: in all the stellar systems, including
the Halo, the HASP fraction is not sensitive to the cut in
amplitude used. This evidence suggests that the Halo sample
incompleteness is far from severe.}.  

We will first try to obtain an upper limit on the Halo fraction
originating in dSph-like systems. For that, the period
distribution of RRab in dSph (panel {\it a} of Fig.~\ref{fig1}) has been rescaled to fill the
maximum possible area under the curves representing the period
distribution for both the inner and outer Halo (red histograms in
Fig.~\ref{fig1}, panels {\it c} and {\it d}). Assuming that the
RRab falling inside the area covered by the red distribution have
been entirely accreted from dSphs, we find that the maximum
contribution of dSph-like systems into the inner and the outer
Halo can not be more than $\sim$50\%. This fraction has to be
cautiously treated. In fact, it is {\it an extreme upper limit}
since the {\it difference\/} between the black and red histograms
consists almost entirely of short-period variables, completely
unlike the observed LMC and Sgr distributions.  Any admixture at
all of these latter two population types to fill in the short
periods would result in far too many halo variables with log P
$\sim-0.22\,$d.  

Even though the above results rely on rough preliminary estimates,
they pose a serious question: {\it Where does the rest (in fact
most) of the Halo mass come from?} There are two main proposed
scenario: {\it i)} from few large and metal--rich stellar systems
LMC or Sgr--like \citep[as suggested by][]{zinn14,tissera14}; {\it
ii)} {\it in situ} stellar formation
\citep[][]{brusadin13,vincenzo14}.  

From the exceptionally complete OGLE III and the new OGLE IV data
for the LMC and the Sgr dwarf respectively, presented in
Section~\ref{hasp}, we have noticed that their RRab populations
share similar properties to that of the Halo in terms of mean
period and sigma, and HASP fraction (Table~\ref{tabHASP}).
We remember here that the application of a one--dimensional
Kolmogorov--Smirnov (KS) test on the LMC, dSphs, GGCs, inner and
outer halo period distributions \citep{stetson14b} strongly
support the evidence that their cumulative distributions are not
drawn from the same parent population (probability $\lsim$0\%).
Interestingly enough, the one--sample KS--test applied to the new
Sgr RRab population does support, with a not negligible
probability 
(10\%\footnote{ 
This result is not affected by possible presence of Sgr RRLs
in the outer Halo. We neglected RRLs with d$_G\gsim30 Kpc$
\citep[][]{zinn14,drake13} and both the mean period 
(<Pab> = 0.576$\pm$0.001 [0.07]) and the correlation 
(10\%) attain similar values.}
), that the outer Halo and Sgr are drawn from the
same distribution. We highlight here that the exceptional
completeness of the huge OGLE LMC sample may partially hide
similarities between the RRab Halo and LMC distributions.  

We performed the same exercise described above, in order to
estimate the fraction of the Halo that may have formed from
systems similar to the LMC or Sgr. We apply a scaling factor to
the LMC and Sgr RRab period distribution in order to match the
largest possible fraction of the halo distribution. We find that
$\sim$80--90\% of the halo may have been formed from this kind of
stellar systems, thus supporting the hypothesis {\it i)}.  
The HASP fraction of the Halo further supports previous
conclusions \citep{venn04,helmi06} that typical dSphs played a
minor role, if any, in its early formation. In this paper we have
provided evidence that HASP RRL are missing in dSph
because these galaxies did not reach a high enough metallicity 
during the time they were able to produce RRLs.
In other words, in their
internal chemical evolution the dSphs achieved a metallicity
[Fe/H]$\sim$--1.5 {\it too recently in the past\/} for stars of a
mass suitable for making RR Lyraes to be currently evolving from
the main sequence.  The LMC, Sgr, and the Halo, in
contrast, achieved these higher metallicities more quickly.  This
gives an indication that the early chemical enrichment histories
of dSphs and more massive stellar systems are dissimilar, in the
sense that chemical enrichment was faster in larger galaxies. 
This is in agreement with the well defined scaling relation
between mass/luminosity and metallicity
\citep{chilingarian11,schroyen13,kirby08} obeyed by dwarf
galaxies, with more massive galaxies being more metal-rich, and
having a broader metallicity distribution.  It follows that the
Halo was made primarily from progenitor galaxies
larger than those that survived to become today's dSphs.  
Future surveys like GAIA will provide a census of a significant fraction of the Halo,
discovering more than 70,000 new RRLs \citep{eyer00}. This is the
required new information to constrain whether the major contributors of the
Halo should be sought in massive dwarf galaxies---LMC
and/or Sgr--like---or in some different formation scenario.  


\acknowledgments
Financial support for this work was provided by FIRB 2013 (RBFR13J716,
PI G. Fiorentino), IAC (grant 310394), ESMS (grant AYA2010-16717) and 
PRIN--MIUR (2010LY5N2T, PI F. Matteucci). We thank an anonymous referee 
for her/his pertinent suggestions. 

\bibliographystyle{apj} 

\end{document}